\documentclass{optica-article}

\journal{opticajournal} 

\articletype{Research Article}

\usepackage{lineno}
\usepackage{xspace}
\usepackage{amsmath}
\usepackage{graphicx}
\usepackage{xcolor}
\usepackage{comment}
\usepackage{multirow}
\usepackage{rotating}
\usepackage{bm}

\newcommand{\etal}{\textit{et al.\@}\xspace}

\newcommand{\Invitro}{\textit{In vitro}\xspace}

\newcommand{\enface}{\textit{en face}\xspace}
\newcommand{\um}{\(\muup\)m\xspace}

\newcommand{\squm}{\(\muup\)m$^2$}
\newcommand{\cbum}{\(\muup\)m$^3$\xspace}
\newcommand{\sqdb}{dB$^2$\xspace}
\newcommand{\invs}{s$^{-1}$\xspace}

\newcommand{\cdeg}{$^\circ$C\xspace}
\newcommand{\cotwo}{CO$_2$\xspace}

\newcommand{\OCDSl}{OCDS$_l$\xspace}

\graphicspath{{./figure}{.}}
\begin{document}

\title{Experimental and numerical investigation of wavelength and resolution dependency of dynamic optical coherence tomography signals}

\author{%
	Shumpei Fujimura\authormark{1},
	Ibrahim Abd El-Sadek\authormark{1,2},
	Rion Morishita\authormark{1},
	Shuichi Makita\authormark{1},
	Atsuko Furukawa\authormark{4},
	Pradipta Mukherjee\authormark{1,3},
	Yiheng Lim\authormark{1},
	Lida Zhu\authormark{1},
	Yunake Feng\authormark{1},
	Thitiya Seesan\authormark{1},
	Satoshi Matsusaka\authormark{4},
	and Yoshiaki Yasuno\authormark{1,*}
}

\address{%
	\authormark{1}Computational Optics Group, University of Tsukuba, Tsukuba, Ibaraki, Japan.\\
	\authormark{2}Department of Physics, Faculty of Science, Damietta University, New Damietta City, Damietta, Egypt.\\
	\authormark{3}Centre for Biomedical Engineering, Indian Institute of Technology Delhi, New Delhi, India.\\
	\authormark{4}Clinical Research and Regional Innovation, Faculty of Medicine, University of Tsukuba, Tsukuba, Ibaraki, Japan.\\
}

\email{\authormark{*}yoshiaki.yasuno@cog-labs.org}

\begin{abstract*} 
The wavelength and system-resolution dependencies of dynamic optical coherence tomography (DOCT) are investigated experimentally and numerically.
Experimental investigations demonstrate significant wavelength dependency for the DOCT values but no resolution dependency.
Numerical simulations were performed using diffusion, random-ballistic motion, and mono-directional flow-based motion models.
Diffusion and random-ballistic motion-based simulations show significant wavelength dependency.
Additionally, small but certain resolution dependency was revealed by these simulations.
Mono-directional flow simulations did not show wavelength dependency, but did demonstrate resolution dependency.
The DOCT value is sensitive to both tissue dynamics and the OCT system specification.
These effects should be considered when interpreting DOCT images.
\end{abstract*}

\section{Introduction}
\Invitro samples and small animals such as spheroid, organoid, and zebrafish are important samples for medical and pharmaceutical research \cite{Hirschhaeuser2010JB, Fei2022BE, Berghmans2005BT}.
Although these samples have thicknesses of only a few millimeters, the lack of non-invasive and label-free microscopic imaging methods hampers effective usage of such samples.

Optical coherence tomography (OCT) is a label-free three-dimensional (3D) imaging modality that offers resolutions of a few micrometers and an image penetration depth of a few millimeters \cite{Huang1991Science, Bouma2009COB, OCT-textbook2015}.
OCT has been widely accepted for use in clinical routines in ophthalmology and cardiology\cite{Everett2020TBiophon, Vignali2014CCR}, and is recently used for microscopic imaging \cite{Izatt1994OL, Huang2019JVE, SELin2021BOE, Ming2022Biosens,KYChen2024SciRep}.

Although OCT-based microscopy can provide the 3D architectures of samples with high resolution, the method only visualizes the structural properties of the sample and offers no sensitivity to the tissue’s functions and activities. 
This lack of functional and activity sensitivity prevents OCT from becoming a standard microscopic imaging tool.
Dynamic OCT (DOCT) \cite{Apelian2016BOE, Ren2024CB} uses a combination of time-sequential OCT signal acquisition and successive signal fluctuation analyses, and may potentially overcome the limited functional and activity sensitivity problems of OCT.

Many types of DOCT algorithm have been presented.
The first type of DOCT algorithm contrasts the tissue activity by using the magnitude of an OCT signal fluctuation, e.g., the variance or the standard deviation \cite{Apelian2016BOE, Oldenburg2012BOE, Oldenburg2015OPG, Thouvenin2017JBO, AbdElSadek2020BOE, Park2021BOE, Scholler2019OE, Morishita2024arXiv}. 
A second type of method uses the speed of OCT signal alternation to form the image contrast.
Some proposed methods use the decay rate of the autocorrelation function of the OCT signal sequence, i.e., the decorrelation speed \cite{Oldenburg2015OPG, AbdElSadek2020BOE}, while others use the dependence of the signal variance on the total sampling time for the OCT signal sequence \cite{Morishita2024arXiv}.
This type of method is believed to be able to visualize the speed of intracellular motions.
The third type of method uses the time spectrum of the OCT signal sequence to generate pseudo-color DOCT image that distinctively contrasts several types of tissues and cells based on their time-frequency properties \cite{Ling2017LSM, Thouvenin2017IOV, Muenter2020OL, Leung2020BOE, Xia2023Optica, Chen2024BOE}.

Among these methods, the authors have used two DOCT algorithms extensively.
The first is the logarithmic intensity variance (LIV) \cite{AbdElSadek2020BOE}, which is a type-1 method that uses the time variance of the dB-scaled OCT signal intensity as the contrast source.
The second is the OCT correlation decay speed (OCDS) \cite{AbdElSadek2020BOE}, which is a type-2 method.
This algorithm computes the slope of the autocorrelation curve of an OCT signal sequence and then uses this slope as the contrast source.
In previous studies, we have simply assumed that the LIV and OCDS values were mainly defined by the intracellular and intratissue dynamics of the sample. 
However, based on an overview of previous LIV- and OCDS-based studies \cite{ElSadek2021BOE, AbdElSadek2023SR, AbdElSadek2024SR, Morishita2023BOE, Mukherjee2021SR, Mukherjee2022BOE, Mukherjee2023SR}, it was suspected that the DOCT values obtained are not only defined by the sample dynamics, but are also strongly influenced by the specifications of the OCT system, including the probe wavelength and the resolutions. 

In this paper, we investigate the dependency of the LIV and OCDS values on both the probe wavelength and the resolution in detail.
To ensure that the investigation is suitably comprehensive, we use two approaches: experimental and numerical approaches.
In the experimental approach, we used two OCT systems that have different wavelengths and resolutions.
In addition, we optionally manipulated the resolution via post-data-acquisition signal processing.
This manipulation method has enabled an accurate experimental investigation solely for the resolution effect by excluding other experimental fluctuating factors.
In the numerical approach, we used a newly developed mathematical model of the tissue dynamics \cite{Yuanke2024SPIE, Yuanke2025arXiv} in combination with a fast simulation method for OCT signal sequences \cite{Morishita2024arXiv} based on a recently proposed OCT imaging model \cite{Tomita2023BOE}.
These investigations clarify the relationships between the DOCT values and the wavelength and resolution of the system.

\section{Experimental investigation}
\subsection{Methods for experimental investigation}
\label{sec:principle}
To investigate the wavelength and resolution dependencies of the LIV and OCDS obtained from biological samples, tumor spheroids were measured using two OCT systems with different wavelengths (840 nm and 1310 nm bands) and resolutions.
To investigate the wavelength dependency when it is not biased by the resolution dependency, we computationally reduced the resolution of one system as it becomes identical to that of the other system.
In contrast, to investigate the resolution dependency, we used the data obtained from one system only, reduced the resolution computationally, and then compared the LIV and OCDS values obtained from the original and resolution-reduced OCT image sequences.

\subsubsection{OCT systems}
\label{sec:OCT system}
Two OCT systems were used in this study.
One is a spectral domain OCT (SD-OCT) with a probe-beam center wavelength of 840 nm, and resolutions of 4.9 \um (lateral, $1/e^2$ width) and 3.8 \um (axial, full-width-half-maximum (FWHM), in tissue) \cite{Morishita2023BOE}.
The other is a Jones-matrix swept-source (SS-) OCT with a probe-beam center wavelength of 1310 nm and resolutions of 18 \um (lateral, $1/e^2$ width) and 14 \um (axial, FWHM, in tissue) \cite{Li2017BOE, Miyazawa2019BOE}.
Although the latter system is polarization sensitive, only a polarization insensitive image, which is the averaged intensity of four polarization channels, was used.
The scanning speeds of both systems are 50,000 A-lines/s.
The details of these systems can be found elsewhere \cite{Morishita2023BOE, Li2017BOE, Miyazawa2019BOE}.
Here after these two systems are denoted as ``high-resolution 840-nm OCT'' and ``low-resolution 1310-nm OCT.''

\subsubsection{DOCT acquisition}
\label{sec:DOCT acquisition}
\begin{figure}
	\centering\includegraphics[width=5cm]{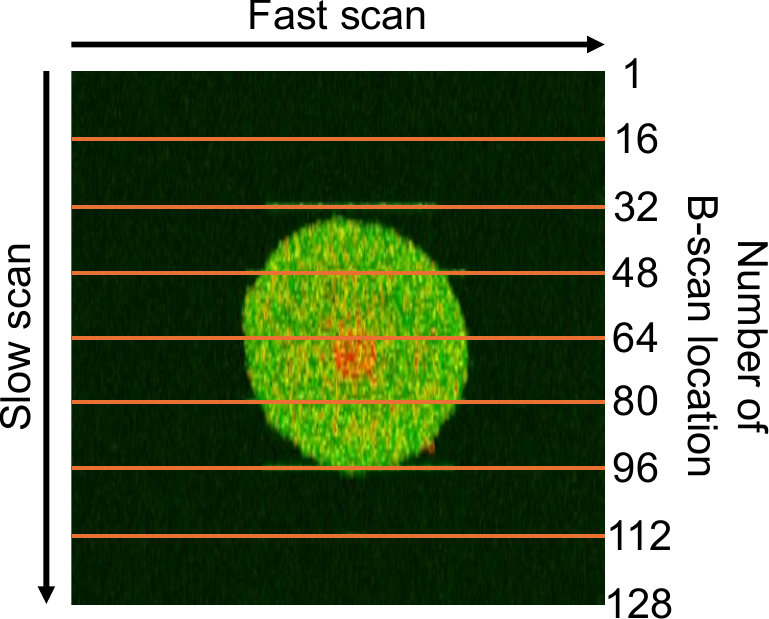}
	\caption{
		Boundaries of eight \enface blocks.
		The \enface field is split into eight blocks, and each block repeated raster-scanned for 32 times.
		The consistency of the OCT phase is not high enough around the block boundaries, and thus the B-scans adjacent to the boundaries were not used in the analyses.
	}
	\label{fig:Boundary issue}
\end{figure}
The three-dimensional (3D) DOCT scanning protocol was used for volumetric DOCT acquisition \cite{ElSadek2021BOE}.
In this protocol, 1-mm $\times$ 1-mm lateral field was divided into eight blocks as schematically illustrated in Fig.\@ \ref{fig:Boundary issue}, and each block consists of 16 B-scan positions.
These blocks were repeatedly raster-scanned for 32 times in 6.55 s.
The frame interval was 204.8 ms and the separation between the first and last frames at a single location was 6.35 s.
The whole OCT volume consisting of 128 B-scan locations was captured in 52.4 s.

Two types of DOCT, i.e., LIV and OCDS were computed from the acquired dataset.
LIV is defined as the variance of the dB-scaled intensity over the 32-frame sequence.
To obtain the OCDS, the auto-correlation function of the OCT time-sequence are computed, the decorrelation slope at a particular delay range of [204.8 ms, 1228.8 ms] was computed, and this slope was used as the OCDS value. 
Note that the OCDS with this particular delay range is denoted as \OCDSl in other publications \cite{ElSadek2021BOE, AbdElSadek2023SR, AbdElSadek2024SR}.
It should be noted that the resolutions of the OCT images were optionally manipulated before computing the LIV and OCDS by the method described in the next section.

\subsubsection{Computational reduction of resolution}
\label{sec:ResReduction}
Both the axial and the lateral resolutions of 840-nm SD-OCT images can be optionally reduced using numerical methods.
Here, we used different methods to manipulate the axial and lateral resolutions.

The lateral resolution was reduced by convolving a spatial Gaussian function with the complex OCT data.
Here, the Gaussian function was applied to the 2D \enface complex OCT image at each depth.
This convolution operation makes the lateral complex point spread function (PSF) as
\begin{equation}
	\mathrm{cPSF}_{\bm{x}} (\bm{x}) =
	\exp\left(-\frac{1}{2}\frac{\bm{x}\cdot\bm{x}}{\sigma_x^2}\right)*
	\exp\left(-\frac{1}{2}\frac{\bm{x}\cdot\bm{x}}{{\sigma'}_x^2}\right)\propto
	\exp\left(-\frac{1}{2}\frac{\bm{x}\cdot\bm{x}}{\sigma_x^2+{\sigma'}_x^2}\right),
	\label{eq:LowResoComplex}
\end{equation}
where $\mathrm{cPSF}_{\bm{x}}$ is the lateral complex PSF, $\bm{x} = (x, y)$ is the lateral 2D position, $\sigma_x$ and the ${\sigma'}_x$ are the width parameters (standard deviations) of the original PSF and the convolved Gaussian function, respectively.
Namely, the first and the second exponential parts are the original lateral PSF and the convolved Gaussian function, respectively.

Since the intensity PSF is the absolute square of the complex PSF, the $1/e^2$-widths of the original and modified intensity PSFs, i.e., the original resolution ($\Delta x$) and the modified (i.e., target) lateral resolution ($\Delta x_t$) are
\begin{equation}
	\label{eq:1/e width of intensity}
	\Delta x = 2\sqrt{2}\sigma_x,
\end{equation}
and 
\begin{equation}
	\label{eq:LowReso 1/e width of intensity}
	\Delta x_t = 2\sqrt{2\left(\sigma_x^2+{\sigma'}_x^2\right)}.
\end{equation}
By combining these two equations, the standard deviation of the spatial Gaussian filter can be defined from the resolutions of the original and target lateral resolutions as
\begin{equation}
	\label{eq:Sigma of filter}
	{\sigma'}_x = \frac{1}{2}\sqrt{\frac{\Delta x_t^2- \Delta x^2}{2}}.
\end{equation}

For axial resolution, the resolution was reduced by numerical spectral apodization using a spectral Gaussian window.
In this case, a spectral Gaussian window $\exp\left[-\frac{1}{2}\left(k/\sigma_s\right)^2\right]$ is applied to the interference spectrum.
Here $k$ does not exactly represents the physical wavenumber but is the Fourier pair of the depth ($z$) and its origin locates at the center of the spectrum.
$\sigma_s$ is the width (standard deviation) of the apodization window.

By this spectral window, the complex PSF along the depth ($z$) is convolved with the Fourier transform of the Gaussian window  $\exp\left[-\frac{1}{2}\left(z/{\sigma'}_z\right)^2\right]$ with ${\sigma'}_z \equiv 1/\sigma_s$, and an equation similar to Eq.\@
(\ref{eq:LowResoComplex}) is given as
\begin{equation}
	\mathrm{cPSF}_z (z)=
	\exp\left(-\frac{1}{2}\frac{z^2}{\sigma_z^2}\right)*
	\exp\left(-\frac{1}{2}\frac{z^2}{{\sigma'}_z^2}\right)\propto
	\exp\left(-\frac{1}{2}\frac{z^2}{\sigma_z^2+{\sigma'}_z^2}\right),
	\label{eq:LowResoComplexZ}
\end{equation}
where $\sigma_z^2$ is the standard-deviation width of the original complex depth PSF.

Unlike the lateral resolution, the axial resolution is defined as the FWHM of the intensity PSF, and hence, the original and modified (i.e., target) axial resolutions ($\Delta z$ and $\Delta z_t$, respectively) can be written using $\sigma_z$, and ${\sigma'}_z$ as
\begin{equation}
	\label{eq:FWHM of intensity}
	\Delta z = 2\sqrt{\log2}{\sigma_z},
\end{equation}
and
\begin{equation}
	\label{eq:LowReso FWHM of intensity}
	\Delta z_t = 2\sqrt{\log2\left(\sigma_z^2+{\sigma'}_z^2\right)}.
\end{equation}

By combining these equations, the spectral apodization window width can be defined from the original and target resolutions as
\begin{equation}
	\label{eq:Sigma of depth filter}
	\sigma_s = {\sigma'}_z^{-1}
	= 2\sqrt{\frac{\log2}{\Delta z_t^2- \Delta z^2}}.
\end{equation}

In our specific implementation, the lateral resolution manipulation was implemented in Python 3.9.17 where the generation of a Gaussian function and the a subsequent convolution were performed using the ``ndimage.gaussian\_filter()'' function of SciPy 1.11.2 library.
The axial resolution manipulation was implemented in LabVIEW, since the spectral apodization is performed in the pipeline of the basic OCT reconstruction process, and this basic OCT reconstruction was implemented in LabVIEW in our specific implementation.

As described previously in Section \ref{sec:DOCT acquisition}, the lateral field was divided into eight blocks in our 3D DOCT scan protocol.
This protocol causes a long acquisition time gap (approximately 6.6 s) between adjacent B-scans at the sub-field boundaries (the horizontal lines  shown in Fig.\@ \ref{fig:Boundary issue}).
Here, the sample dynamics deteriorates the consistency of the complex OCT signal over the boundary, and it deteriorates the accuracy of the lateral complex convolution operation around the boundary. 
Therefore, the B-scans adjacent to the boundary were excluded from the subsequent analyses.

\subsubsection{Sample and measurement protocol}
\label{sec:Sample and measurement protocol}
Ten human breast adenocarcinoma spheroids (MCF-7 cell line) were used in this study.
The details of the culture protocols are described in elsewhere \cite{AbdElSadek2020BOE}.
In short, 1,000 human breast-derived tumor cells were seeded into each well of a 96-well ultra-low-density plate and cultured for 5 days to form the spheroids.

Before the measurements start, the spheroids were stored in a portable incubator (Cell Box Ground 2.0, Cellbox Solutions, Germany) in 5\% \cotwo concentration and at a temperature of 37.0 \cdeg.
Two spheroids were extracted from the incubator at the same time, placed in different Petri dishes, and one spheroid was measured first by the high-resolution 840-nm OCT and then by the low-resolution 1310-nm OCT.
At the same time, the other spheroid was measured first by the 1310-nm OCT and then by 840-nm OCT.
And hence, the effect of the measurement order in statistical data analyses will be reasonably minimized.
The total measurement time of each spheroid was 13.3 $\pm$ 5.4  minutes.
According to the previous longitudinal DOCT imaging study of spheroids \cite{ElSadek2021BOE}, this relatively short measurement time does not affect the viability of the spheroids.

\subsubsection{Study design}
\label{sec:Study design}
To investigate the wavelength dependency, the resolutions of high-resolution 840-nm OCT were reduced to match those of low-resolution 1310-nm OCT before computing the LIV and OCDS by the method described in Section \ref{sec:ResReduction}.
To investigate the resolution dependencies, only the dataset acquired by the high-resolution 840-nm OCT was used.
To investigate the axial resolution dependency, the axial resolution was numerically modified into 3.8, 7.6, 11.4, and 14 \um, while the lateral resolution was not modified.
To investigate the lateral resolution dependency, the lateral resolution was numerically modified into 4.9, 9.8, 14.7, and 18 \um as keeping the original axial resolution.

The mean LIV and OCDS were computed over the spheroid region but the B-scans adjacent to the block boundary were excluded (see the last paragraph of Section \ref{sec:ResReduction}).

The wavelength dependency of the mean LIV and OCDS values were examined by paired t-test.
On the other hand, the resolution dependency was examined first by applying Kruskal-Wallis test among the four resolutions, and if significance was found, paired t-test was performed between each pair of resolutions.
The SciPy functions of “scipy.stats.ttest\_rel()” and “scipy.stats.kruskal()” were used to perform the paired t-test and the Kruskal-Wallis test, respectively.

\subsection{Results}
\subsubsection{Wavelength dependency}
\label{sec:WaveDepeExperiment}
\begin{figure}
	\centering\includegraphics[width=8.7cm]{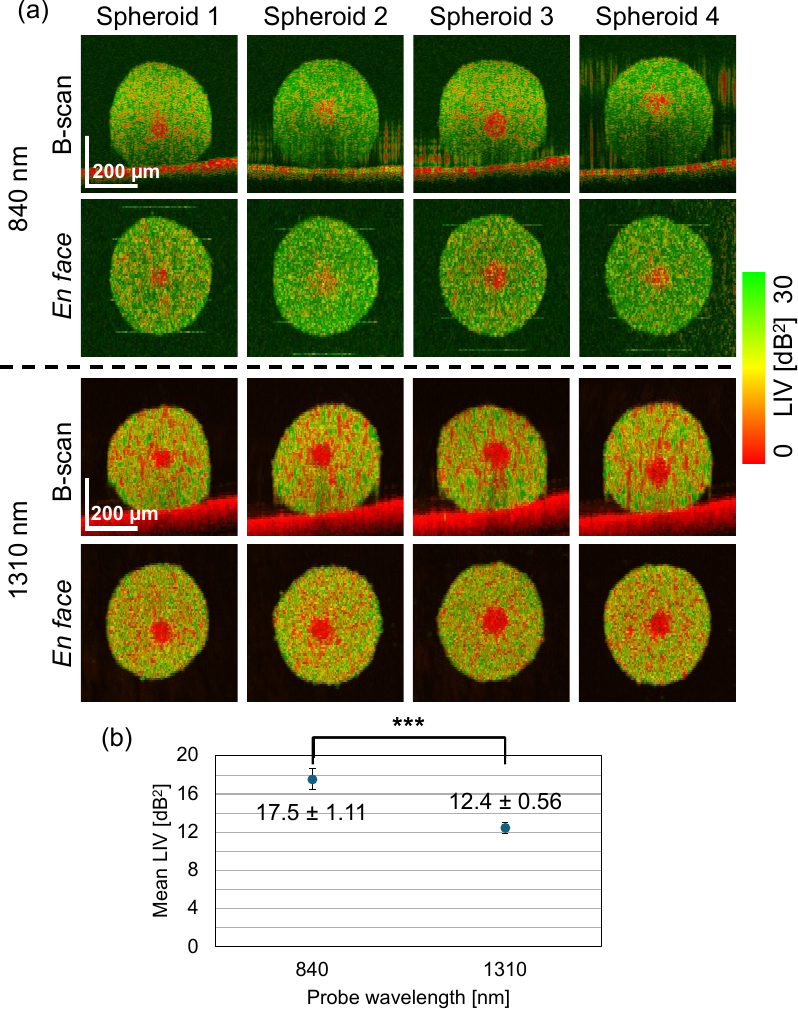}
	\caption{%
		Wavelength dependency of LIV.
		(a) LIV images of spheroids acquired with short (840-nm) and long (1310-nm) OCT systems.
		Both the lateral and axial resolutions of 840-om OCT were computationally reduced to match those of 1310-nm OCT.
		(b) Mean LIV value of whole spheroid volume.
		The dots indicate the means of mean LIV and whiskers indicate the standard deviations among the four spheroids.
		The shorter wavelength gave significantly higher LIV with statistical significance (p < 0.001, as indicated by ***).
	}
	\label{fig:LIVWaveDepe}
\end{figure}
Figure \ref{fig:LIVWaveDepe}(a) shows the cross-sectional and \enface LIV images of four spheroids measured with 840-nm and 1310-nm OCT systems.
Here, both the lateral and axial resolutions of 840-om OCT were computationally reduced to match those of 1310-nm OCT.
For all spheroids, the 840-nm images show higher LIV (i.e., green) than the 1310-nm images.
To quantitatively validate this difference, the mean LIV of the entire spheroid volume was computed and plotted in Fig.\@ \ref{fig:LIVWaveDepe}(b).
The dots and the whiskers indicate the mean among the four spheroids and its standard deviation, respectively.
(This representation also stands for latter plots.)
As expected from the image observation, the shorter wavelength gave a significantly higher mean LIV than 1310-nm OCT (5.09-\sqdb higher, p < 0.001).

\begin{figure}
	\centering\includegraphics[width=8.7cm]{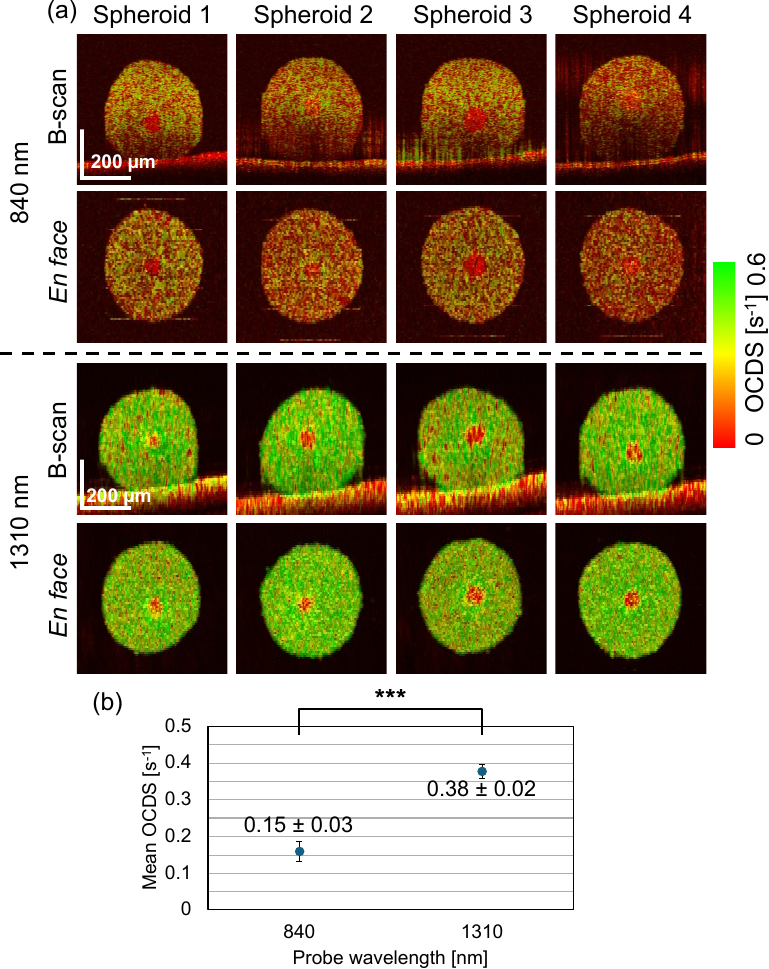}
	\caption{%
		Wavelength dependency of OCDS.
		(a) OCDS images of spheroids acquired with short (840-nm) and long (1310-nm) OCT systems.
		Both the lateral and axial resolutions of 840-om OCT were computationally reduced to match those of 1310-nm OCT.
		(b) The mean OCDS value of whole spheroid volume.
		The dots indicate the means of the mean OCDS and the whiskers indicate the standard deviations among the four spheroids.
		The shorter wavelength gave significantly lower OCDS with statistical significance (p < 0.001, indicated by ***).
	}
	\label{fig:OCDSWaveDepe}
\end{figure}
The OCDS images also showed a clear wavelength dependency as shown in Figure \ref{fig:OCDSWaveDepe}(a).
In the case of OCDS, not like the LIV, the 1310-nm images show higher OCDS (green) than the 840-nm images.
A quantitative comparison of the mean OCDS for entire spheroid volume validated this subjective observation.
Namely, the mean of the mean OCDS of the four spheroids was 0.22-\invs higher for the longer wavelength (1310 nm) than for the shorter wavelength (840 nm) [Fig.\@ \ref{fig:OCDSWaveDepe}(b)] with statistical significance (p < 0.001).

These experiments confirmed that both LIV and OCDS have significant dependencies on the probe wavelength.

\subsubsection{Axial-resolution dependency}
\label{sec:AxialDepeExperiment}
\begin{figure}
	\centering\includegraphics[width=13cm]{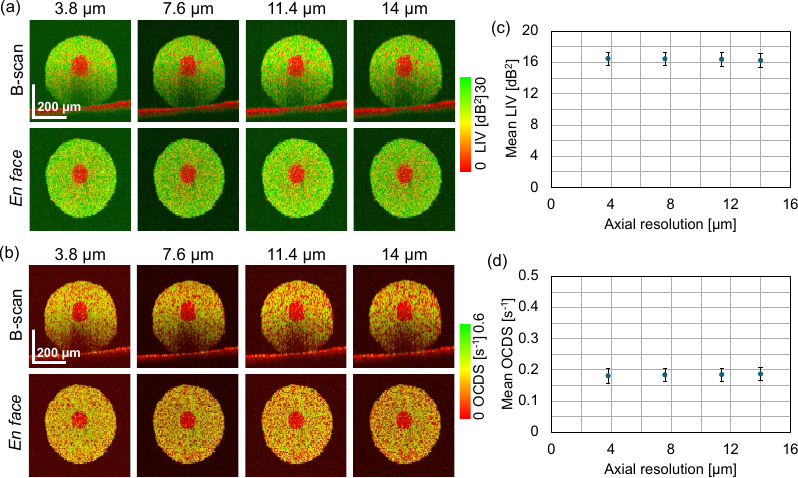}
	\caption{%
		Axial-resolution dependency of LIV and OCDS.
		(a, b) LIV and OCDS images of a representative tumor spheroid at four different axial resolutions.
		All images were generated from an identical sequence of complex OCT image obtained from 840-nm OCT but with the computational-resolution modification.
		(c, d) Means of (c) mean LIV and (d) mean OCDS.
		Each of the mean LIV and OCDS were computed from the entire volume of each spheroid, and the means and the standard deviations of the four spheroids were plotted.
		Both LIV and OCDS showed no significant axial-resolution dependency.
	}
	\label{fig:ExpAxialDepend}
\end{figure}
Figure \ref{fig:ExpAxialDepend}(a) and (b) show the cross-sectional and \enface (a) LIV and (b) OCDS images of a representative spheroid with four different axial resolutions.
Here, all images were obtained from the same dataset obtained by the high-resolution 840-nm OCT, and a computational modification of the axial resolution was applied to the complex OCT.
Both the LIV and OCDS images do not show any clear differences among the resolutions.
The results were also validated by statistical test.
Namely, the plots of LIV [Fig.\@ \ref{fig:ExpAxialDepend}(c)] and OCDS [Fig.\@ \ref{fig:ExpAxialDepend}(d)] are smooth and quite constant, and no statistically significant differences were found among the resolutions (p = 0.89 for LIV and p = 0.95 for OCDS).

\subsubsection{Lateral resolution dependency}
\label{sec:LateralDepeExperiment}
\begin{figure}
	\centering\includegraphics[width=13cm]{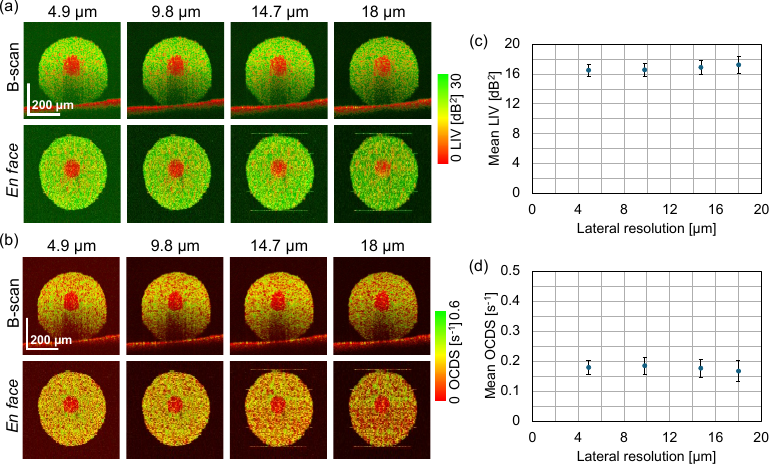}
	\caption{%
		Lateral-resolution dependency of LIV and OCDS.
		(a, b) LIV and OCDS images of a representative tumor spheroid at four different lateral resolutions.
		All images were obtained from high-resolution 840-nm OCT but with the computational-resolution modification.
		(c, d) Means of (c) mean LIV and (d) OCDS.
		Each mean LIV and OCDS was computed from the entire volume of each spheroid, and the means and the standard deviations of the four spheroids were plotted.
		Both LIV and OCDS showed no significant lateral-resolution dependency.
	}
	\label{fig:ExpLateralDepend}
\end{figure}
LIV and OCDS images of a representative spheroid with different lateral resolutions are shown in Fig.\@ \ref{fig:ExpLateralDepend}(a) and (b), respectively.
Similar to the axial-resolution dependency evaluation, all images were obtained from the same high-resolution 840-nm OCT datasets, and the lateral resolutions were computationally manipulated after the data acquisition.

It was subjectively observed that the lateral resolution causes no marked differences between the LIV and OCDS images.
The mean LIV and OCDS plots [shown in Fig.\@ \ref{fig:ExpLateralDepend}(c) and (d), respectively] support this absence of a lateral-resolution dependency. 
Here, the plots show the means and standard deviations of the whole-spheroid-volume mean LIV and mean OCDS, and they do not exhibit significant differences among the lateral resolutions (p = 0.39 for LIV and p = 0.66 for OCDS).

\section{Methods for numerical investigation}
\label{sec:simulationMethod}
\subsection{Methods for numerical investigation}
\subsubsection{Scatterer model}
\label{sec:Scatterer model}
The numerical validation was performed using our recently released open-source DOCT simulation framework \cite{DoctLibrary, Yuanke2025arXiv}.
In the numerical validation, we investigate the wavelength and resolution dependencies of both LIV and OCDS obtained from scatterers moving with various motion types and speeds.
In our simulations, the dynamic samples are mathematically modeled by a modified version of dispersed-scatterer model (DSM) \cite{Tomita2023BOE}. 
The original DSM represents the sample as a set of infinitely small scatterers that are randomly dispersed in a spatially slowly varying refractive index medium.
We simplified this model by assuming that the spatially slowly varying refractive index is uniform (i.e., constant) over the analytic area and the refractive indexes of all scatterers are identical.
In addition, we extended the DSM to account for the sample dynamics as follows.

Three scatterer motion types were considered in the numerical emulation of the intracellular activities.
The first is diffusion where the scatterers' displacements over two sampling time points are stochastically determined by following a zero-mean normal distribution as
\begin{equation}
	\label{eq:Diffusion model}
	\begin{split}
		&x_{j,i+1} = x_{j,i} + \mathcal{N}\left(\mu = 0, \sigma^2 = 2D \Delta t \right),\\
		&y_{j,i+1} = y_{j,i} + \mathcal{N}\left(\mu = 0, \sigma^2 = 2D \Delta t \right),\\
		&z_{j,i+1} = z_{j,i} + \mathcal{N}\left(\mu = 0, \sigma^2 = 2D \Delta t \right).
	\end{split}
\end{equation}
where $(x_{j,i}, y_{j,i}, z_{j,i})$ represents the 3D position of the $j$-th scatterer at the $i$-th time point $t_i$, and $\Delta t \equiv t_{i+1}- ,t_i$ is the time increment in the numerical simulations.
$\mathcal{N}\left(\mu = 0, \sigma^2 = 2D \Delta t \right)$ is a random variable that follows the zero-mean normal distribution with a variance of $\sigma^2 = 2D \Delta t$.
$D$ is a predefined diffusion coefficient.

The second type of motion is random-ballistic motion, which is expressed as
\begin{equation}
	\label{eq:Random and flow model}
	\begin{split}
		&x_{j,i+1} = x_{j,i} + v \Delta t \sin\varphi_j\cos\theta_j,\\
		&y_{j,i+1} = y_{j,i} + v \Delta t \sin\varphi_j\sin\theta_j,\\
		&z_{j,i+1} = z_{j,i} + v \Delta t \cos\varphi_j,\\
	\end{split}
\end{equation}
where $v$ is the scatterer speed.
$\varphi_j$ and $\theta_j$ are the azimuth and inclination angles of the $j$-th scatterer, respectively.
In this simulation, $v$ is constant for all scatterers and for all simulation time, but $\varphi_j$ and $\theta_j$ are randomly selected for each scatterer while following a constant distribution over $(-\pi, \pi]$.

The third type of motion is mono-directional flow, in which all scatterers ballistically move with the same speed and direction.
Namely, the scatterers follow the same equation as the random-ballistic case [Eq.\@ \ref{eq:Random and flow model}], but both $\varphi_j$ and $\theta_j$ are identical for all scatterers.

\subsubsection{DOCT signal generation}
\label{sec:Simulated DOCT signal}
To obtain the DOCT signal, a sequence of OCT signals are obtained using the method described in Ref.\@ \cite{Morishita2024arXiv}. 
Briefly, the complex OCT signal at a particular point in the image is obtained as a function of time as
\begin{equation}
	\label{eq:single complex OCT}
		E(t_i) \propto \sum_{j=0}^{N_s-1} \textcolor{teal}{b_j \exp\left[i\frac{2\pi}{\lambda}2z_j(t) + i\phi_z \right]} \\
		\textcolor{purple}{\exp\left[
			-\frac{1}{2}\left(\frac{x_{j,i}^2 + y_{j,i}^2}{\sigma_x^2}\right) 
			-\frac{1}{2}\left(\frac{z_{j,i}}{\sigma_z}\right)^2\right]},\\
\end{equation}
where$(x_{j,i}, y_{j,i}, z_{j,i})$ is the 3D position of the $j$-th scatterer at the $i$-th time point $t_i$.
The green part of the equation is the phasor contribution to the complex OCT signal from each scatterer.
$\phi_z$ is a constant phase offset defined by the path length offset of the OCT interferometer.
$b_j$ is the complex reflectivity of the $j$-th scatterer, where we assume that $b_j$ is constant for all scatterers and is given as $b_j \equiv \mathrm{constant}$, for simplicity.
The red part of the equation is the complex Gaussian PSF of OCT, where $\sigma_x$, and $\sigma_z$ denote the standard-deviation widths of the PSF along the lateral and axial directions, respectively. 
Here we assume that the lateral resolutions are isotropic for $x$ and $y$ directions.

Finally, the complex OCT signal is yielded by collecting the phasor contributions from all scatterers, which are weighted by the complex PSF over the sample as shown in the equation.
$N_s$ in the equation is the number of scatterers in the sample.

Finally, the OCT signal intensity is obtained as
\begin{equation}
	\label{eq:intensity}
	I(t) = E(t)E^{*}(t),
\end{equation}
where the superscript $^{*}$ denotes the complex conjugate.

In our specific implementation of the simulation, the size of the 3D area of the simulation was set to be sufficiently large to account for the size of the PSF, the motions of the scatterers, and the total time to be simulated (see Section 3.1.3 of Ref.\@ \cite{Morishita2024arXiv} for details).
The scatterers are seeded within this 3D area, i.e., the initial position of each scatterer is randomly selected within this area.
For each time increment, the positions of the scatterers are updated as to follow one of the motion models, i.e., the diffusion, random-ballistic motion, or mono-directional flow [Eqs. (\ref{eq:Diffusion model}) or (\ref{eq:Random and flow model})].
The OCT signal at each time point is computed by Eqs.\@ (\ref{eq:single complex OCT}) and (\ref{eq:intensity}).
The LIV and OCDS are then computed from the simulated OCT time sequence. 
The specific parameters and simulation conditions are described in the next subsection.

\subsection{Study design}
\label{sec:Simulaltion study design}
The wavelength and resolution dependencies of both LIV and OCDS were investigated by numerical simulations.
In the simulations, the OCT system parameters were set to match those of the OCT systems used in the experimental validation described in Section \ref{sec:OCT system}, 
i.e., two wavelengths of 840 and 1310 nm, and two resolution sets including high-resolution (4.9 \um for lateral and 3.8 \um for axial) and low resolution (18 \um for lateral and 14 \um for axial), where the lateral and axial resolutions are defined as the $1/e^2$ width and the FWHM, respectively.
It should be noted that we simulated all combinations of the wavelength and resolution sets including high-resolution 1310-nm OCT and low-resolution 840-nm OCT, although some of which do not correspond to the real OCT devices used in the experimental study.

The time points of the simulation were selected to be consistent with those in the experiments described in Section \ref{sec:DOCT acquisition}.
Namely, we simulated the time-sequential OCT signal consisting of 32 time points with an OCT frame interval of 204.8 ms, and the total time window is 6.35 s. 
The time range of the correlation curve fitting of OCDS is also identical to the experiment, i.e., [204.8 ms, 1228.8 ms].

The simulations were performed for the three types of motion individually, where the motion types are diffusion, random-ballistic motion, and mono-directional flow. 
For diffusion, the diffusion coefficient $D$ were selected from the range of [0, 15 \squm/s]. 
For random ballistic and mono-directional flow motions, the velocity of scatterers is in the range of [0, 6.0 \um/s]. 
The maximum-diffusion coefficient of 15 \squm/s and -speed of 6.0 \um/s were selected so that the maximum displacement of the scatterer during the whole simulation duration becomes sufficiently larger than the largest resolution among all the resolution sets, i.e., 18 \um.
Specifically, the maximum scatterer displacement in the 6.3-s time window become 41.2 \um (for diffusion) and 37.8 \um (for both random ballistic and mono-directional flow), where the maximum displacement of the diffusion case was defined as $3\sigma = 3\sqrt{2D \Delta t}$.

The scatterer density was set to 0.055 scatterers/\cbum, and was selected according to the scatterer density measured from the tumor spheroids by a neural-network-based scatterer-density estimator \cite{Seesan2021BOE, Seesan2024BOE}. 

For each set of parameters, 500 trials were performed with different randomly seeded scatterers and the mean LIV and OCDS over the trials are used as the results.

\subsection{Results}
\subsubsection{Wavelength dependency}
\label{sec:WaveDepeSimu}
\begin{figure}
	\centering\includegraphics[width=11.8cm]{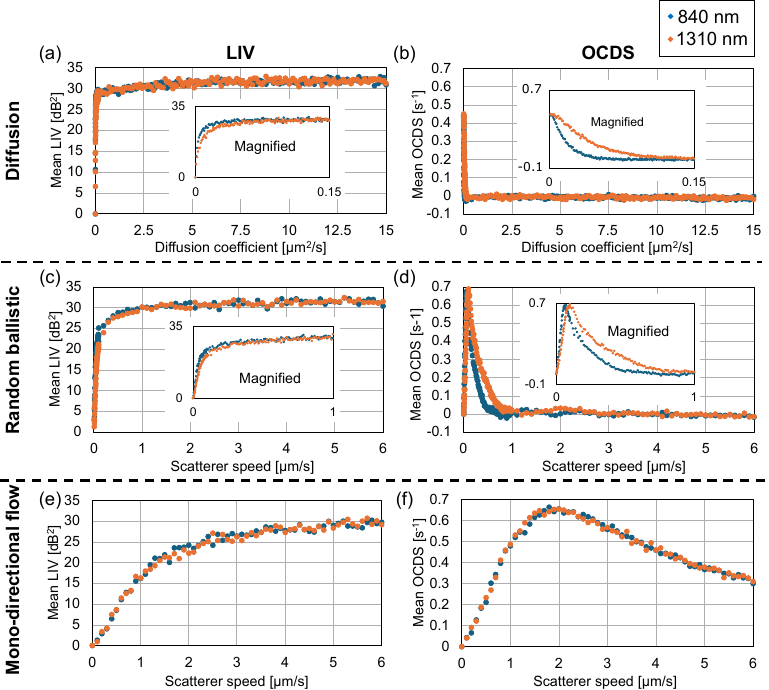}
	\caption{%
		Wavelength dependency of LIV and OCDS in (a,b) diffusion, (c,d) random-ballistic motion, and (e,f) mono-directional flow.
		The plot color indicates the wavelength as shown in the legends (blue for 840 nm and orange for 1310 nm).
		The resolutions for both wavelengths are 4.9 \um in the lateral and 3.8 \um in the axial.
		Wavelength dependency was observed for both the diffusion and random-ballistic motions, but was not observed for the mono-directional flow.
	}
	\label{fig:WaveDepeSimuResults}
\end{figure}
Figure \ref{fig:WaveDepeSimuResults} shows the simulated LIV and OCDS at two different wavelengths (blue for 840 nm and orange for 1310 nm) for diffusion (first row), random-ballistic motion (second row), and mono-directional flow (third row).
The LIV and OCDS are plotted against the diffusion coefficient or the speed of the scatterers.
Here, the resolutions for both wavelengths are identical to those of the high-resolution set, i.e, 4.9 \um for lateral and 3.8 \um for axial.
The direction of the mono-directional flow was set to be 45 degrees from the depth axis.

The diffusion and random-ballistic motion cases show wavelength dependency in both LIV and OCDS especially at small diffusion coefficients and scatterer speeds [see the magnified insets in Fig.\@ \ref{fig:WaveDepeSimuResults}(a-d)], while these differences become negligible for large diffusion coefficients and scatterer speeds.
On the other hand, for the mono-directional flow, both the LIV and OCDS do not show marked wavelength dependency over the whole speed range [Fig.\@ \ref{fig:WaveDepeSimuResults}(e,f)].

In summary, wavelength dependency was found for both the diffusion and the random-ballistic motion, but it was not observed for the mono-directional flow.
The mechanism of this difference among the motion types is discussed in detail in Section \ref{sec:InterWave}.

\subsubsection{Resolution dependency}
\begin{figure}
	\centering\includegraphics[width=11.8cm]{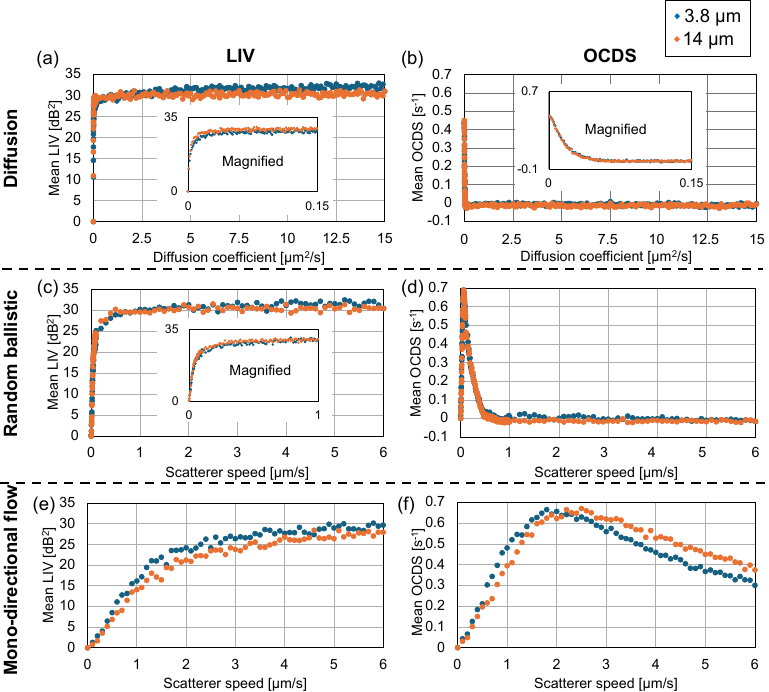}
	\caption{%
		Axial resolution dependency of LIV and OCDS for the cases of (a,b) diffusion, (c,d) random-ballistic motion, and (e,f) mono-directional flow.
		The point color indicates the axial resolution as shown in the legend (blue for 3.8 \um and orange for 14 \um).
		The lateral resolution and the wavelength are set to 4.9 \um, and 840 nm, respectively.
		For both diffusion and random-ballistic motion, the LIV showed tiny but certain resolution dependency, and these dependencies are also altered by the diffusion coefficient or the scatterer speed.
		The mono-directional flow showed significant axial resolution dependencies for both LIV and OCDS.
	}
	\label{fig:AxialDepeSimuResults}
\end{figure}
Figure \ref{fig:AxialDepeSimuResults} shows the simulated LIV and OCDS for the three motion types at two axial resolutions (blue for 3.8 \um and orange for 14 \um).
For both cases of these axial resolutions, the lateral resolution was set to 4.9 \um, and the wavelength was set to 840 nm.
The direction of the mono-directional flow was set to be 45 degrees from the depth axis.

For the diffusion case, the low resolution (orange) gives a slightly higher LIV than the high resolution (blue) for small diffusion coefficients [see Fig.\@ \ref{fig:AxialDepeSimuResults}(a) and its magnified inset].
However, as the diffusion coefficient becomes large, such as more than 1.2 \squm/s, the low resolution LIV (orange) becomes lower than that of high resolution (blue).
Not like the LIV, the OCDS of the diffusion does not show an axial-resolution dependency [Fig.\@ \ref{fig:AxialDepeSimuResults}(b)] .

The LIV of the random-ballistic motion shows similar properties to the diffusion case, namely, the low resolution gives a higher LIV than high resolution when the scatterer speed is low, and this relationship is inverted when the scatterer speed becomes high [Fig.\@ \ref{fig:AxialDepeSimuResults}(c) and its magnified inset].
The OCDS of the random-ballistic motion with high axial resolution is slightly higher than that of low axial resolution at around the scatterer speed range of 0.5 to 5 \um/s [Fig.\@ \ref{fig:AxialDepeSimuResults}(d)].

For the mono-directional flow, both the LIV and the OCDS show marked axial-resolution dependencies [Fig.\@ \ref{fig:AxialDepeSimuResults}(e,f)].

\begin{figure}
	\centering\includegraphics[width=11.8cm]{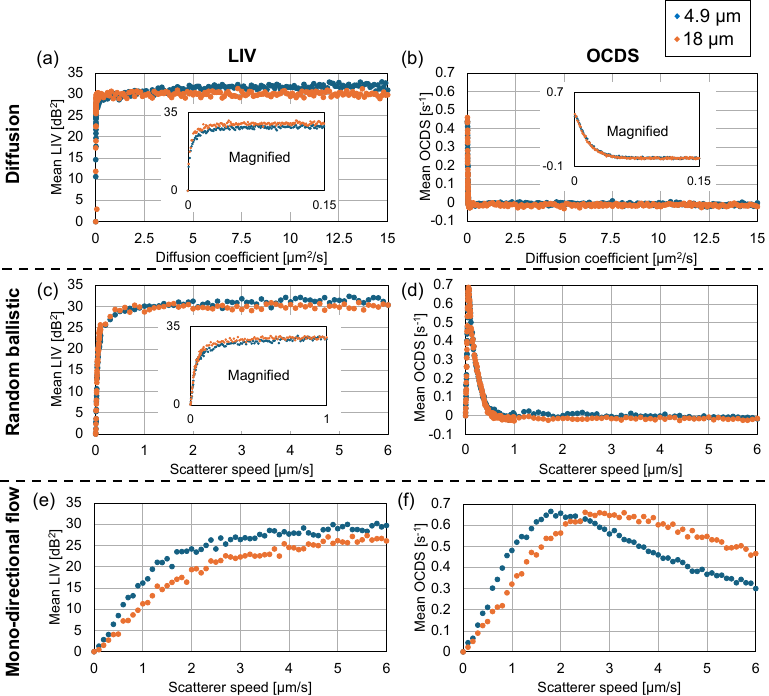}
	\caption{%
		Lateral resolution dependency of LIV and OCDS for (a,b) diffusion, (c,d) random-ballistic motion, and  (e,f) mono-directional flow.
		The point color indicates the lateral resolution as shown in the legends (blue for 4.9 \um and orange for 18 \um).
		The axial resolution and the wavelength for both lateral resolutions are 3.8 \um, and 840 nm, respectively.
		Results similar to the axial resolution dependency analyses were observed.
	}
	\label{fig:LateralDepeSimuResults}
\end{figure}
Figure \ref{fig:LateralDepeSimuResults} summarizes the simulated LIV and OCDS for the three motion types at two lateral resolutions (blue for 4.9 \um and orange for 18 \um).
In both cases, the axial resolution was set to 3.8 \um and the wavelength was set to 840 nm.
The direction of the mono-directional flow was set to be 45 degrees from the depth axis.
Each plot shows similar result to the case of axial-resolution dependency.

The mechanism of the resolution dependency is discussed in detail Section \ref{sec:InterReso}.

\section{Discussion}
\subsection{Mechanism of wavelength dependency}
\label{sec:InterWave}
Both the experimental and numerical investigations showed strong wavelength dependency, except for the numerical investigation of the mono-directional flow.
Since the intracellular motion of the spheroid cannot be a mono-directional flow, the experimental and numerical results are consistent.

The mechanism of the wavelength dependency of the diffusion and random-ballistic motion can be explained as follows.
If the scatterer motion is diffusion or random ballistic, the relative positions among the scatterers change over time.
Particularly, when the relative depth positions of the scatterers change, the mutual phases among the scattered lights from scatterers are altered, and this change causes alternation of interference intensity over time.
And this alteration is captured by DOCT.
Since the mutual phase of the scattered lights is proportional to the depth distance of the scatterers and inversely proportional to the wavelength, the alternation of the interference intensity is dependent on the wavelength.
This explains the wavelength dependency of both the LIV and the OCDS.

In addition, shorter the wavelength will give larger the phase change, and hence larger the OCT intensity fluctuation.
The larger LIV values at the shorter wavelength found in both the experiment and the simulation are consistent with this depiction.

\subsection{Mechanism of resolution dependency}
\label{sec:InterReso}
Although the experiments did not show marked resolution dependency, the simulations of the LIV for both diffusion and random-ballistic motion showed small but certain resolution dependency [Figs.\@ \ref{fig:AxialDepeSimuResults} and \ref{fig:LateralDepeSimuResults}].
This discrepancy between the experiment and the simulations may be because the dependency was too small to be observed by the experiment.

\newcommand{\propo}[1]{{\textbf{P{[#1]}}\xspace}}
\newcommand{\propoderive}[1]{{$\leftarrow$ \propo{#1}\xspace}}
A remarkable finding of the simulations of the diffusion and the random-ballistic motion was that the lower resolution gives higher LIV if the diffusion coefficient or the scatterer speed is small, while the higher resolution gives higher LIV if the diffusion coefficient or the scatterer speed is large.
This biphasic property can be explained by the following logic.
In the following descriptive logic diagram, \propo{a} represents Proposition a, and \propoderive{b, c} at the begging of the proposition means that the proposition was derived from Propositions b and c.
The interrelation among the propositions are also graphically summarized in Fig.\@ \ref{fig:logcDiagram}.

\begin{figure}
	\centering\includegraphics[width=8cm]{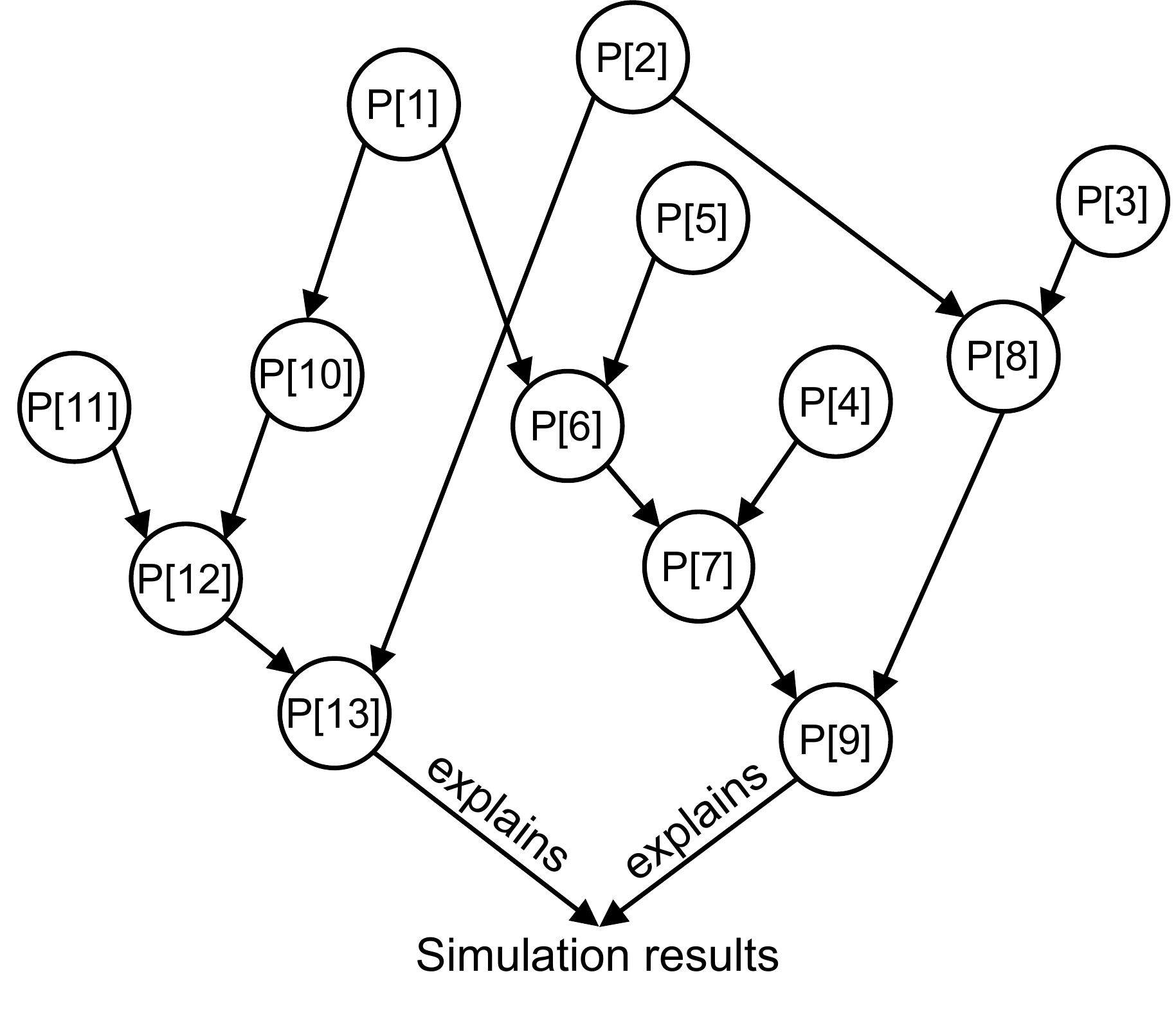}
	\caption{
		The interrelation among the propositions to explain the resolution dependency of LIV found in the simulation results.
		\propo{a} means Proposition a, and \propo{a} $\rightarrow$ \propo{b} means Proposition b is supported by Proposition a.
		\label{fig:logcDiagram}
	}
\end{figure}
\propo{1}: The scatterer displacements during the OCT-sequence acquisition are statistically proportional to the diffusion coefficient or the scatterer speed, which are collectively denoted as ``motion parameter,'' later on.
Namely, the horizontal axes of the plots shown in Figs.\@ \ref{fig:AxialDepeSimuResults} and \ref{fig:LateralDepeSimuResults} are proportional to the displacements of the scatterers.

\propo{2}: Each resolution configuration has a different effective number of scatterers (ENS), which is the number of scatterers within the resolution volume.
In our simulation, the ENS is 5.02 if the (axial resolution, and lateral resolution) are (high, high), 18.5 for (low, high), and 67.7 for (high, low).
Note that (high, high) is of the blue plots (high-resolution cases) of both axial- and lateral-resolution dependency analyses [Figs.\@ \ref{fig:AxialDepeSimuResults} and \ref{fig:LateralDepeSimuResults}].

\propo{3}: 
According to Hillman \etal \cite{Hillman2006OL}, the spatial speckle contrast is very high if the ENS is less than around 2, and the contrast rapidly decreases as the ENS increases, and becomes almost constant and low for ENS more than around 10 (see Fig.\@ 3 of Ref.\@ \cite{Hillman2006OL}).

\propo{4}:
The LIV can be considered to monotonically relate to the temporal speckle contrast.

\propo{5}:
The temporal speckle contrast approaches to the spatial speckle contrast as the displacement of the scatterers becomes large.

\propo{6} \propoderive{1, 5}:
The temporal speckle contrast approaches to the spatial speckle contrast as the motion parameter becomes large.

\propo{7} \propoderive{6, 4}:
The LIV at a very large motion parameter reaches a maximum that is defined by the spatial speckle contrast.

\propo{8} \propoderive{2, 3}:
The spatial speckle contrast of the (high, high)-resolution configuration must be higher than that of both the (high, low)- and (low, high)-resolution configurations.

\propo{9} \propoderive{7, 8}:
The LIV of the high resolution configurations must be higher than that of low resolution at the large motion parameter.
This logical inference agrees with the simulation results.

\propo{10} \propoderive{1}:
The scatterer displacement must be very small at the very left part of the plots.

\propo{11}:
If the ENS is very small, such as smaller than one, the very small displacement does not significantly alter the OCT intensity. 
Note that, if ENS is one, the scattered lights are not mutually interfered and speckle is not formed.
And hence, the temporal speckle contrast is far smaller than the spatial speckle contrast.
If the ENS becomes 2, the scattered light from the scatterers interfered to each other, and a small mutual displacement alters the OCT intensity.
As the ENS increases further, the phase of the interference signal more easily alter, and hence the OCT intensity more easily alters.
And hence, even if the displacement amount of each scatterers were small, the temporal speckle contrast becomes higher when the ENS is larger.
In other words, the temporal speckle contrast more quickly approaches to its maximum, i.e., the spatial speckle contrast.

\propo{12} \propoderive{10, 11}:
The plotted LIV more quickly reaches to its flat top in Figs.\@ \ref{fig:AxialDepeSimuResults}(a, c) and \ref{fig:LateralDepeSimuResults}(a, c) for larger ENS values.

\propo{13} \propoderive{2, 12}:
The plotted LIV more quickly reaches to its flat top for low-resolution configurations.
This explains why the lower resolution gave the higher LIV when the motion parameter is small in the numerical simulations.

\textbf{Conclusion:}
\propo{9} and \propo{13} explain the biphasic property of LIV shown in Figs.\@ \ref{fig:AxialDepeSimuResults}(a, c) and \ref{fig:LateralDepeSimuResults}(a, c).

For the mono-directional flow simulations, the high resolution gave a higher LIV than the low resolution for all scatterer speeds [Figs.\@ \ref{fig:AxialDepeSimuResults}(e) and \ref{fig:LateralDepeSimuResults}(e)].
This is because the flow does not alter the mutual distances among the scatterers, and hence the spatial speckle patterns does not alters but only shifts, where the solid speckle pattern shifts at the same speed with the scatterers.
This eliminates the effects of \propo{11} and \propo{12}, namely, the saturation speed of the curve is not necessarily increased by ENS and hence by the resolution volume.
On the other hand, since the spatial speckle size is statistically similar to the resolution, the signal at a point in the image more rapidly alters for high-resolution even if the scatterer speed is the same.
Namely, the temporal speckle contrast and also the LIV more rapidly saturate over the scatterer speed for higher resolutions.

\section{Conclusion and future perspective}
The wavelength and resolution dependencies of DOCT (i.e., LIV and OCDS) were investigated by experiments and simulations of diffusion, random-ballistic motion, and mono-directional flow.
Both the experiment and simulation showed significant wavelength dependency of LIV and OCDS for the diffusion and random-ballistic motion.
Although no marked resolution dependency was observed in the experiment, the simulation revealed tiny but characteristic resolution dependency for the diffusion and random-ballistic motion.
This tiny resolution dependency was theoretically explained by the mechanism of LIV.
The simulation also revealed that the LIV and OCDS values of mono-directional flow are also affected by the resolutions.

These results implicate that DOCT measurement is not only sensitive to the tissue dynamics but also affected by the specifications of the OCT system. 
This fact should be considered when we observe and interpret DOCT images in application studies.

The numerical analysis revealed that different motion types cause different characteristics of the wavelength and resolution dependencies.
It may be used to classify motion types from DOCT values if proper inference methods are designed in future work.
In addition, by exploiting the numerically revealed relationship between the DOCT values and the motion parameters (i.e., the diffusion coefficient or the scatterer speed), methods to estimate motion parameters from DOCT can possible be designed.
A preliminary version of this motion parameter estimation method has been demonstrated \cite{Fujimura2025BiOS}.
More detailed and accurate numerical investigations and sophisticated estimation principles, such as deep learning, may enable more accurate and comprehensive parameter estimation, and hence, enables quantitative DOCT measurement.

\section*{Funding}
Core Research for Evolutional Science and Technology (JPMJCR2105); 
Japan Society for the Promotion of Science (21H01836, 22F22355, 22KF0058, 22K04962, 24KJ0510);
Japan Science and Technology Agency (JPMJFS2106). 

\section*{Disclosures.}
Fujimura, El-Sadek, Morishita, Makita, Mukherjee, Lim, Zhu, Feng, Seesan, Yasuno: Sky Technology(F), Nikon(F), Kao Corp.(F), Topcon(F), Panasonic(F), Santec (F), Nidek (F);
Furukawa, Matsusaka: None.

Fujimura, Lim, and Zhu are currently employed by Hitachi High-Tech Science Corp., Ainnovi, and Santec Holdings, respectively.

\section*{Data availability.}
The data that support the findings of this study are available from the corresponding author upon reasonable request.

\bibliography{Bibliography}

\end{document}